\documentclass[twocolumn,aps,prb]{revtex4}
\usepackage{epsfig,graphicx}
\begin{document}

\title {\bf On the structure of the Si(103) surface}

\author{Cristian V. Ciobanu\footnote{Author to whom correspondence should be
addressed; electronic mail: cciobanu@mines.edu (C.V. Ciobanu)}}
 \affiliation{Division of Engineering,
Colorado School of Mines, Golden, Colorado 80401}
\author{Feng-Chuan Chuang}
\affiliation{Department of Physics, National Sun Yat-Sen
University, Kaohsiung, 804, Taiwan}
\author{Damon E. Lytle}
\affiliation{Department of Metallurgical and Materials
Engineering, Colorado School of Mines, Golden, Colorado 80401}

\begin{abstract}
Although (103) is a stable nominal orientation for both silicon and germanium,
experimental observations revealed that in the case of silicon this surface remains
disordered on an atomic scale even after careful annealing.
We report here a set of low-energy reconstruction models corresponding to
$1\times 2$, $2\times 2$, and $1\times 4$ periodicities, and propose
that the observed disorder stems from the presence of several coexisting reconstructions with
different morphologies and nearly equal surface energies. The reconstructions found
also suggest that the models previously reported in the literature for the (103)
orientation have very high surface energies and are thus unlikely to be experimentally observed.
\end{abstract}
\maketitle

In recent years, the high-index semiconductor surfaces have steadily
gained in technological and fundamental importance.  From the technological standpoint,
these surfaces have clear potential serve as templates for growing linear arrays of
nanostructures because they can have a stepped or grooved
morphology with characteristic lengths in the nanoscale
regime. Some high-index orientations, however, are nominally
flat and are often observed to be the facets of the quantum dots
formed during heteroepitaxial growth. It is the case, for example,
of the (105) facets that bound the pyramidal islands obtained in the
Ge/Si(001) system.\cite{mo} To date, a number of high-index Si and Ge
surfaces have been discovered to be stable,\cite{highindexcatalogSiGe}
i.e. they do not facet into other orientations.

Among the stable surfaces of Si and Ge that so far have received
very little attention from a theoretical perspective are
Si(103) and Ge(103). Despite the fact that they have the same orientation,
experiments indicate that they have very different atomic structure and morphology.\cite{thermalstab-notsame-asGe1x4}
Ge(103) exhibits two-dimensional atomic ordering with a clear periodic
pattern,\cite{Ge1x4-germany,
Ge1x4-china} while Si(103) remains rough and disordered on the atomic scale even
after careful annealing.\cite{SS-rough103-105,thermalstab-notsame-asGe1x4} This remarkable difference
between Si(103) and Ge(103) is, in itself, a fundamentally interesting problem.
Still, because the Si(103) surface is atomically rough and thus
very difficult to tackle, so far there has not been sufficient motivation for
performing extensive structure studies on this surface. This situation changes with the
discovery\cite{hut-103-facet} of the (103) facetted pyramids
that appear during the Si overgrowth of the Ge/Si(001) quantum dots. Motivated by the
recent experiments of Wu {\em et al.},\cite{hut-103-facet} we have
set out to find atomic structure models for Si(103). Based on these models, we
suggest that the rough and disordered aspect of Si(103) is due to the
coexistence of several reconstructions of similar energies and different
bonding topologies. Furthermore, the reconstructions presented here provide
evidence that the Ge(103)-$1\times 4$ models previously reported\cite{Ge1x4-germany,Ge1x4-china}
have too high surface energies to be confirmed in experiments.

\begin{figure}
  \begin{center}
   \includegraphics[width=6.5cm]{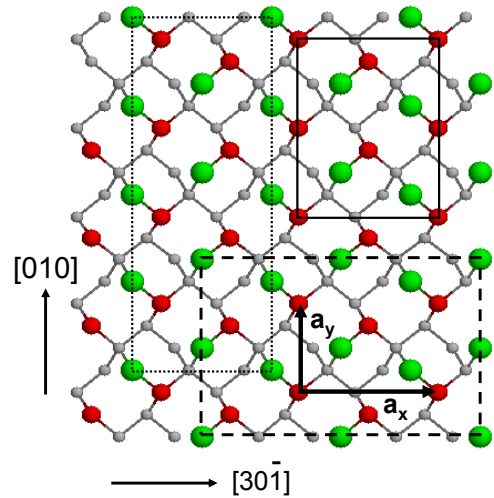}
\end{center}
\caption{(Color online) Top view of the bulk truncated Si(103) surface.
The larger (green) atoms have two dangling bonds, the intermediate-sized (red) ones
have one dangling bond, and the small gray atoms are four coordinated.
The unit vectors of the $1\times1$ unreconstructed primitive cell are $\mathbf{a_x}=a{\sqrt{2.5}}\mathbf{e_x}$
and $\mathbf{a_y}=a\mathbf{e_y}$, where $a=5.431$\AA \ is the lattice constant of Si, and $\mathbf{e_x}$
and $\mathbf{e_y}$ are the unit vectors along $[30{\overline 1}]$ and [010], respectively.
The rectangles show the unit cells for the $1\times 2$ (solid line),
the $2\times 2$ (dash line), and the $1\times 4$ (dotted line)  reconstructions.} \label{geom}
\end{figure}

\begin{figure}
  \begin{center}
   \includegraphics[width=8.7cm]{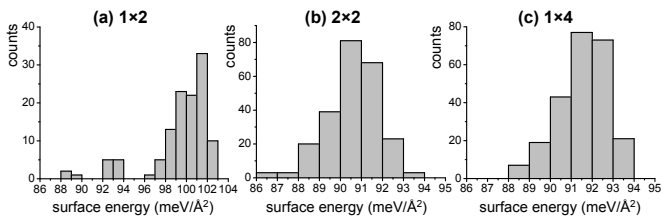}
\end{center}
\caption{Histograms of the surface energies retrieved by the genetic algorithm for the (a)
Si(103)-$1\times 2$, (b) Si(103)-$2\times 2$, and (c) Si(103)-$1\times 4$ reconstructions.}\label{histograms}
\end{figure}

The structural models for the Si(103) orientation were determined using
a genetic algorithm optimization\cite{105-ga} coupled with the
Lenosky {\em et al.} highly-optimized empirical potential (HOEP)\cite{hoep} model of atomic interactions.
We have considered three sizes of the computational cell, $1\times 2$, $2\times 2$, and $1\times 4$,
which are shown in Fig.~\ref{geom}. The algorithm selects structures
based on the surface energy $\gamma$, and starts with a 'genetic pool' of $p=30$
initially random configurations of the top 5\AA\ of the surface slabs.
The genetic pool evolves through cross-over operations which combine
portions of two randomly chosen pool members (parents) to create a new
structure (child). The child structure is relaxed and retained in the pool
if its surface energy is sufficiently low.\cite{105-ga} The optimization is performed for each of the possible
numbers of atoms (kept constant) that yield distinct global minima of a given surface slab.
Since there are four atoms in an $1\times 2$ layer (Fig.~\ref{geom}), we have performed four runs for this
supercell size and eight runs for each of the other two sizes, $2\times 2$
and $1\times 4$. The surface energies corresponding to the 600 model
reconstructions retrieved are organized in the histograms shown in Fig.~\ref{histograms}.

\begin{figure}[t]
  \begin{center}
   \includegraphics[width=8.5cm]{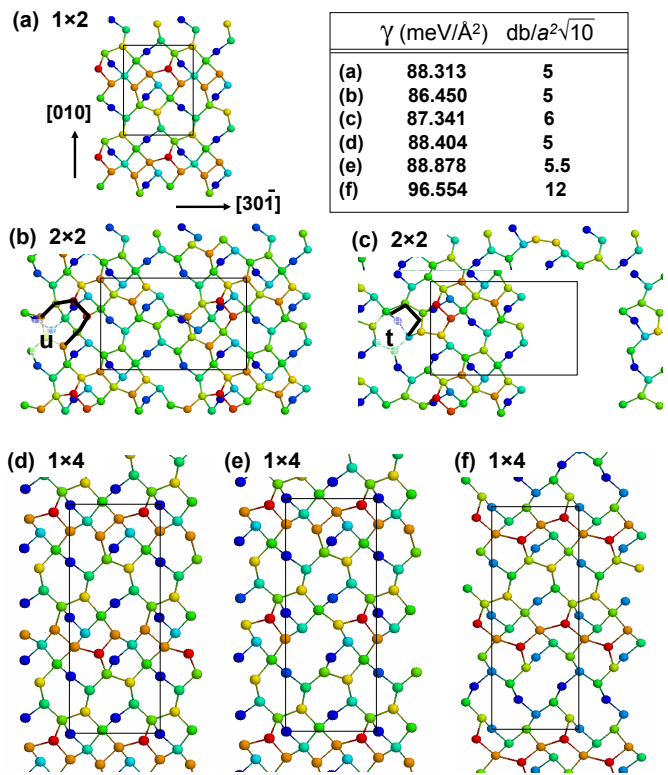}
\end{center}
\caption{Model reconstructions (top views) for the Si(103) surface (a)--(f).
Structures (a) through (e) have been obtained in this work, while panel (f)
shows the model previously proposed the (103) orientation.\cite{Ge1x4-germany,Ge1x4-china}
Atoms are colored according to their coordinates along [103], from red
(highest position) to blue (lowest position in the slab shown); the periodic
cell is marked by a rectangle in each case. The table(inset) shows
the surface energy $\gamma$ of models (a)--(f) calculated using
the Lenosky {\em et al.}\cite{hoep} potential, and their number of dangling
bonds (db) per $1\times 2$ unit area.}
\label{103reconstructions}
\end{figure}

To analyze the Si(103) models (Fig.~\ref{histograms}), we note that recent
studies of high-index Si surfaces suggest that the correct
(i.e. experimentally confirmed) structure either has the
lowest HOEP surface energy [e.g., Si(105) in Ref.~\onlinecite{105-ga}]
or it has a surface energy that most likely lies within
3--4 meV/\AA$^2$ from the lowest HOEP surface energy value [as in the case of
Si(114) and Si(337)].\cite{114-ga, 337-ga}
Therefore, in order to identify good Si(103) reconstructions we focus
on a surface energy {\em range} that includes most of the thermodynamically favorable structures,
i.e. 86 meV/\AA$^2< \gamma <$~89 meV/\AA$^2$ (refer to Fig.~\ref{histograms}).
In this range, there are 35 models across the three periodicities considered (Fig.~\ref{geom}).
Of these models, 32 are distinct in the sense that the large period structures
($1\times 4$ and $2\times 2$) can not be reduced to the
repetition of a single $1\times 2$ model.

From the 32 distinct structures, we have identified a few pairs of
configurations that exhibit minor differences such as bonds relaxing to sightly
different local minima but otherwise making up the same topology at the surface.
More notably, there are also groups of nearly degenerate reconstructions
with markedly different atomic bonding but with nearly equal
(and low) surface energies. Some of these reconstructions are depicted
in panels (a)--(e) of Fig.~\ref{103reconstructions}. We have
found that the atomic scale features that appear frequently on most of
the favorable reconstructions (not only those shown in Fig.~\ref{103reconstructions})
are the dimers and the rebonded-atoms, which would be expected for stepped Si(001)
surfaces. Dimers and rebonded atoms occur in a wide variety of relative
configurations for any of the low-energy Si(103) reconstructions.
Interestingly, the dimer-rebonded atom configuration that is solely responsible
for the lowest energy structure of Si(105)\cite{105-exp-compression, kds} is also
encountered on Si(103); this configuration is made up of two rebonded atoms that
``bridge" at the base of a dimer to form a shape that resembles somewhat the
letter {\bf \em  u}.\cite{cereda-highsteps-ss} Figure~\ref{103reconstructions} shows
one such {\bf \em u} motif marked in black in panel (b), which can readily be spotted
in the other panels as well. Another known motif that appears (though not as
frequently as the {\bf \em u}) on the low-energy Si(103) reconstructions is the tetramer,\cite{tetramer} denoted
by {\bf \em t} in Fig.~\ref{103reconstructions}(c).

The similarity between the best Si(103) model found here [Fig.~\ref{103reconstructions}(b)]
and the single-height rebonded (SR) model \cite{105-exp-compression} for Si(105) is
quite striking, as they have two {\bf \em u} motifs in their respective unit cells
and nearly equal density of dangling bonds, i.e. 1.58 db/$a^2$ for Si(103) vs. 1.57 db/$a^2$ for SR.
Since the unit cells of Si(105)-$1\times 2$ and Si(103)-$2\times 2$
have different sizes, the best Si(103) model allows for an efficient
arrangement of its motifs at the cost of introducing additional surface stress.
Therefore, the resulting lowest surface energy for Si(103), 86.45~meV/\AA$^2$,
is higher than the surface energy of the the SR model,\cite{105-ga} 82.20~meV/\AA$^2$.

The surface stress associated with low-energy Si(103) structures is tensile, because
most of the bonds are stretched in order to achieve a low dangling bond density.
On the other hand, a very large a number of dangling bonds per area
increases the surface energy even though the atoms at the surface would
have significantly more room to relax. This is the case of the reconstruction
proposed originally for the Ge(103)-$1\times 4$ surface,\cite{Ge1x4-germany, Ge1x4-china} and shown
here in Fig.~\ref{103reconstructions}(f) after scaling to the lattice constant
of Si and relaxation at the HOEP level. The surface energy
of the model in  Fig.~\ref{103reconstructions}(f) is
96.55~meV/\AA$^2$, clearly larger than the surface energy of {\em any} of the 240
structures accounted for in Fig.~\ref{histograms}(c). Even though the main focus
of this paper is not on Ge(103), we were intrigued by finding such a high
surface energy for model (f) so we recalculated the surface energies of
all $1\times 4$ models using an empirical potential for Ge.\cite{tersoff}
We have found a surface energy of 91.67 meV/\AA$^2$ for model (f) with
the Tersoff potential,\cite{tersoff} while the surface energies of all
other $1\times 4$ structures (scaled to the lattice constant of Ge)
ranged between 85.94 meV/\AA$^2$ and 95.02 meV/\AA$^2$. This finding
suggests that a re-evaluation of the accepted Ge(103)-$1\times 4$
model\cite{Ge1x4-germany, Ge1x4-china} may be warranted in the future.

We conclude with a short discussion of the physical implications of
having a large number of low-energy reconstructions available for
the Si(103) surface. The existence of multiple models with similar
surface energies but with very different topologies and
different spatial periodicities suggests that it is possible
for such models to {\em coexist} on the Si(103) orientation, a proposal
which has recently been made for the case of Si(105) as well.\cite{Si105degen}
Indeed, experiments to date\cite{SS-rough103-105} show that both Si(103) and
Si(105) are atomically rough and exhibit no discernable two-dimensional
periodicity even after careful annealing. The proposal that
several structural patterns can coexist on the same nominal orientation
would have little value if any two models placed next to one
another on the (103) surface were to give rise to domain boundaries
with very high formation energies. However, we have found that {\em different} $1\times 2$ models do indeed appear
next to one another without substantially increasing the surface energy of the
reconstructions with larger unit cells: refer, for example, to
Fig.~\ref{103reconstructions}(e), in which the $1\times 2$ model (a) occupies
the upper half of the $1\times 4$ cell. Since there exists a vast array
of energetically favorable motifs made of dimers and rebonded atoms,
entropy considerations also support the idea of various structural
patterns coexisting on the Si(103) surface.

In summary, we have used a genetic algorithm to find a large set of reconstructions
for Si(103), and proposed that the atomic scale roughness experimentally observed
for this surface is due to the coexistence of several nearly degenerate structural models
with different bonding topologies and surface periodicities but with similar surface
energies. By analyzing the Si(103) models, we have found that the low-energy
(103) reconstructions largely display the same atomic-scale motifs (combinations
of dimers and rebonded atoms) as Si(105),\cite{Si105degen} which has lead us to believe
that the physical origin of the observed\cite{SS-rough103-105} disorder is
the same for both Si(103) and Si(105). In the case of Si(105), the
structural degeneracy is lifted upon applying compressive strain\cite{Si105degen} or
through the heteroepitaxial deposition of Ge.\cite{105-exp-compression}
For Si(103) it was shown that low coverages of indium can result in
the emergence of a preferred reconstruction pattern.\cite{indium}
The possibility to remove the degeneracy and create a periodic pattern on Si(103)
by epitaxially depositing Ge at low coverage has not been investigated.\cite{hige}
If such experiments were to be performed, the calculations presented here
predict that the most likely model to emerge is the that in Fig.~\ref{103reconstructions}(b),
which is similar to the SR model that emerges upon deposition of Ge on Si(105).\cite{105-exp-compression}
Upon comparing the structures retrieved by the genetic algorithm with the existing model\cite{Ge1x4-china, {Ge1x4-germany}} for the Ge(103) surface,
we have found that the latter has a density of dangling bonds that is 2.4 times
larger than that of the best (103) models. The models presented here\cite{available} can, we hope, play an important role in
revisiting the currently accepted structure of Ge(103), as well as in
explaining the (103)-facetted islands\cite{hut-103-facet} that appear upon Si
capping of the Ge/Si(001) quantum dots.

{\em Acknowledgments.} CVC gratefully acknowledges the support of
the National Center for Supercomputing Applications through Grant
No. DMR-050031. FCC is supported by the National Science Council
of Taiwan under Grant No. NSC95-2112-M110-022.


\begin{thebibliography}{99}
\bibitem{available} Due to space limitations, Fig.~\ref{103reconstructions} only shows a few reconstructions,
but any (103) models can be made available upon request.
\end{thebibliography}
\end{document}